\documentclass[aps,prd,preprint,groupedaddress]{revtex4-1}
\usepackage[dvips]{graphicx}
\begin{document}

\title{Neutrino and anti-neutrino transport in accretion disks}
\author{Zhen Pan and Ye-Fei Yuan}
\affiliation{Key Laboratory for Research in Galaxies and Cosmology CAS, Department of Astronomy, University of Science and Technology of China, Hefei, Anhui 230026, China}

\begin{abstract}
We numerically solve the one dimensional Boltzmann equation of the neutrino and
anti-neutrino transport in accretion disks and obtain the fully energy dependent and direction dependent neutrino and anti-neutrino emitting  spectra, under condition that the distribution of the mass density,temperature and chemical components
are given. Then, we apply the
resulting neutrino and anti-neutrino emitting spectra to calculate the corresponding annihilation rate of neutrino
pairs above the neutrino dominated accretion disk and find that the released energy resulting from
the annihilation of neutrino pairs can not provide sufficient energy for the most
energetic short gamma ray bursts whose isotropic luminosity can be as high as $10^{52}$ ergs/s unless the
high temperature zone where the temperature is beyond $10$ MeV can stretch over
200 km in the disk. We also compare the resulting luminosity of neutrinos and anti-neutrinos with 
the results from the two commonly used approximate treatment of 
the neutrino and anti-neutrino luminosity: the Fermi-Dirac black body limit and a simplified model of neutrino transport,
i.e., the gray body model,
and find that both of them overestimate the neutrino/anti-neutrino luminosity and their annihilation rate greatly.
Additionally, as did in Sawyer (2003), we also check the validity of the two stream approximation, and find that
it is a good approximation to high accuracy.
\end{abstract}

\maketitle

\section{Introduction}
Short Gamma ray bursts (SGRBs) is one of the most energetic phenomena in nature releasing energy as much as $10^{52}$ ergs in less than 2 seconds \cite{Piran2004}\cite{Zhang2004}\cite{Nakar2007}. Until now the central engine of SGRBs is still an open question though numerous models have been proposed for decades and one of the most commonly accepted model is the 
neutrino dominate accretion flows (NDAFs) involving a hyper-accreting black hole with mass accretion rate about $ 0.1\sim 10$ M$_{\rm {sun}}$/s.
The characteristic mass density is about $ 10^9 \sim 10^{12}$ g/cm$^3$ and the characteristic temperature is about $10^{10}\sim 10^{11}$ K in
the inner part of NDAFs. Therefore, photons are almost completely trapped and only neutrinos/anti-neutrinos can escape from the disk carrying away the gravitational energy of the accreted gas (\cite{Narayan1992}\cite{Popham1999}\cite{DiMatteo2002}\cite{Narayan2001}\cite{Kohri2002}\cite{Lee2005}\cite{Janiuk2007}\cite{Liu2008}). 
Under some extreme conditions with high accretion rate ($> 1$ M$_{\rm {sun}}$/s), even the neutrino opacity
in the disk cannot be neglected, either. When the inner part of the disk is not transparent to neutrinos, \citet{DiMatteo2002} investigated the
neutrino transport in the disk under two main approximations,
the first one is that they applied the energy-averaged cross-sections of neutrino-matter interaction, which is similar to the Rosseland approximation for photons; the second one is the two stream approximation
of the angle dependence of the neutrino distribution.
\citet{Sawyer2003} checked the above two approximations
by solving the full Boltzmann equation directly, his results have shown that the two stream approximation is good enough, but the
energy-averaged approximation is only accurate to order.

Obviously, it is vitally important to solve the neutrino transport problem if we want to obtain a more precise and more realistic neutrino emitting spectrum, i.e., the energy dependent and direction dependent spectrum, which sensitively determines the finial annihilation rate of neutrino pairs. Unfortunately, few works focusing on neutrino transport and neutrino spectrum in NDAFs have been done and the two common approximate treatment of the neutrinos/anti-neutrinos luminosity in the previous works 
are as follows:
first, the neutrino emission is integrated over the volume of the disk neglecting the absorption in the 
neutrino transparent cases and assuming the neutrino spectrum to be the Fermi-Dirac black body limit in 
the neutrino opaque cases: $f(p)=1/(\exp(p/kT)+1)$ (we denote the assumption as Fermi black body spectrum 
for short in the following sections), where $p$ is the energy of neutrinos/anti-neutrinos and $T$ is the local temperature of the disk \cite{Popham1999};
second, a simplified neutrino transport model for both neutrino transparent and opaque cases \cite{DiMatteo2002}, 
i.e., a gray body model for neutrinos/anti-neutrinos spectrum ($f(p)=b/(\exp(p/kT)+1)$, where $b$ is a function of neutrino opacity) is applied, please also see Eq.(14) for reference \cite{Janiuk2007}.
Meanwhile, the neutrino transport problem has been investigated in the atmosphere of neutron stars and in the spherically symmetric supernova.
In the atmosphere of neutron stars, it is supposed that the energy flux and lepton number flux are fixed as boundary conditions and are conserved through the atmosphere (\citet{Schinder1982a}\cite{Schinder1982}), the result shows that the only effect of the atmosphere is to soften the neutrino and anti-neutrino spectra because the neutrino cross section is roughly proportional to the square of the neutrino energy, and therefore the high energy neutrinos are much easier to be absorbed.
Neutrino transport problem in spherically symmetrical supernovas has also been investigated by \citet{Burrows2000} whose result shows that the spectra of neutrinos and anti-neutrinos are neither in the form of the Fermi black body spectrum nor the 
gray body model spectrum. Though the atmosphere of neutron stars and  supernova are not the same as NDAFs, their results and the results of \citet{Sawyer2003} strongly imply that the spectra in the form of
the Fermi black body spectrum or the gray body model spectrum are not guaranteed in arbitrary situation.

Although the neutrino transport in accretion disks was investigated by
solving the full Boltzmann equation directly by \citet{Sawyer2003}, the
main purpose of \citet{Sawyer2003} is to check the validity of two stream approximation and
the energy-averaged approximation, while the author did not give the explicit neutrino spectrum.
In Ref.\cite{Sawyer2003}, the author also made a simplification that the driving chemical potential $\mu_{\rm{eq}}=\mu_{\rm_{e}} + \mu_{\rm{p}} - \mu_{\rm{n}}$ ($\mu_{\rm{e}},\mu_{\rm{p}},\mu_{\rm_{n}}$ is the chemical potential of electron, proton and neutron respectively) of neutrinos is much smaller than the temperature in disks and can be neglected. However, this simplification will smear the
difference between neutrinos and anti-neutrinos which is rather important in determining the resulting neutrino and anti-neutrino spectra and the final annihilation rate of neutrino pairs.

In this paper, we study the spectra of neutrinos and anti-neutrinos by solving the full one dimensional Boltzmann Equation in an infinite and homogeneous disk on the equator plane of a central compact object. All the physical quantities vary in the vertical direction and are symmetrically distributed about the equator of the disk $z=0$. Therefore, the boundary conditions for neutrino distribution function $f(\mathbf p,z)$ can be written as $f(\mathbf p,0)=f(\mathbf {-p},0)$ and $f(\mathbf p,H)=0 $ for $\mathbf p\cdot \hat z <0$ ,where $\mathbf p$ is the momentum of neutrinos and $H$ is the upper boundary of the disk.
In \S II, the Boltzmann equation for neutrinos and the absorption/emission coefficient of neutrinos are presented.
Our numerical methods for solving the Boltzmann equation are briefly given in \S III. 
In \S IV, given the conditions of the disk, we calculate the emitting spectra of both
neutrinos and anti-neutrinos, which are used to calculate their annihilation rate
above the disk in \S V. As did in Ref. \cite{Sawyer2003}, we also check the validity
of two stream approximation in \S VI. Finally, our conclusions and discussions are summarized
in \S VII. 

\section{Boltzmann Equation}
Following the notation of \citet{Sawyer2003},  we define $\mu=\cos(\theta)$ for the up moving neutrinos and  $\mu=-\cos(\theta)$ for the down-moving ones where $\theta$ is the angle of neutrino moving direction to the vertical direction of disk. We also define the distribution function of the up-moving and down-moving neutrinos to be $f_+(z,p,\mu)$ and $f_-(z,p,\mu)$
respectively, where $z$ is vertical coordinate and $p$ is the energy of neutrinos. For the up-moving neutrinos \cite{Sawyer2003}\cite{Schinder1982}\cite{Burrows2006}, their distribution
function is determined by
\begin{equation}
\mu \frac{\partial f_+(z,p,\mu)}{\partial z}=\lambda_a \left[f^{\rm{eq}}(T(z),\mu_{\rm{eq}},p)-f_+(z,p,\mu)\right]+\lambda_s\left[-f_+(z,p,\mu)+\frac{1}{2} \int_0^1 d\mu f_-(z,p,\mu)+f_+(z,p,\mu)\right]
\end{equation}
and for the down-moving ones, their distribution function is determined by
\begin{equation}
\mu \frac{\partial f_-(z,p,\mu)}{\partial z}=-\lambda_a \left[f^{\rm{eq}}(T(z),\mu_{\rm{eq}},p)-f_-(z,p,\mu)\right]-\lambda_s\left[-f_-(z,p,\mu)+\frac{1}{2} \int_0^1 d\mu f_-(z,p,\mu)+f_+(z,p,\mu)\right]
\end{equation}
where $\lambda_a$ is the absorption coefficient, $\lambda_s$ is the isotropic elastic scattering coefficient, in this work, we only include the effects of the
isotropic and elastic scattering in the scattering term, and here
$f^{\rm{eq}}=1/(\exp{((p-\mu_{\rm{eq}})/{kT})}+1)$ for neutrinos, $f^{\rm{eq}}=1/(\exp{((p+\mu_{\rm{eq}})/{kT})}+1)$ for anti-neutrinos where $\mu_{\rm{eq}}=\mu_{\rm_{e}} + \mu_{\rm{p}} - \mu_{\rm{n}}$,  and $\mu_{\rm{e}},\mu_{\rm{p}},\mu_{\rm_{n}}$ is the chemical potential of electron,proton and neutron,respectively.

Specifically, for the neutrino process $\nu_e + n \leftrightarrow e^- + p$, the absorption
coefficient for neutrinos is shown as follow,
\begin{equation}
\lambda_a=n_{\rm{n}}\sigma\frac{(1-f_{\rm{e}})(1-f_{\rm{p}})}{1-f_{\nu_{\rm{e}}}^{\rm{eq}}},
\end{equation}
where  $\sigma=\sigma_0[(1+3g_{\rm{A}}^2)/4]((p+\Delta)/m_{\rm{e}})^2 \sqrt{1-(m_{\rm{e}}/(p+\Delta))^2}(1+1.1p/m_{\rm{n}})$ , $n_{\rm{n}}$ is the number density of neutron and $(1-f_{\rm{e}})(1-f_{\rm{p}})/(1-f_{\rm{\nu_e}}^{\rm{eq}})$ is the final state blocking and the stimulated absorption correction,
 here $f_{\rm{e}},f_{\rm{p}}$ is the Fermi distribution function of electron and proton,
respectively. $\sigma_0$ is the characteristic neutrino cross section ($\sigma_0=1.705 \times 10^{-44}$ cm$^2$), $\Delta$ is the mass gap between neutron and proton ($\Delta=m_{\rm{n}}-m_{\rm{p}}$=1.29 MeV), and $g_{\rm{A}}$ is the axial-vector coupling constant ($g_{\rm{A}}\sim-1.23$).

For the anti-neutrino process $\overline{\nu}_e + p \leftrightarrow e^+ + n$, the resulting
absorption coefficient is written as,
\begin{equation}
\lambda_a=n_{\rm{p}}\sigma\frac{(1-f_{\rm{e^+}})(1-f_{\rm{n}})}{1-f_{\overline{\nu}_{\rm{e}}}^{\rm{eq}}},
\end{equation}
where $\sigma=\sigma_0[(1+3g_{\rm{A}}^2)/4]((p-\Delta)/m_{\rm{e}})^2 \sqrt{1-(m_{\rm{e}}/(p-\Delta))^2}(1-7.1p/m_{\rm{n}})$, $n_{\rm{p}}$ is the number density of proton and $(1-f_{\rm{e^+}})(1-f_{\rm{n}})/(1-f_{\overline{\nu}_{\rm{e}}})$ is the final state blocking and the stimulated absorption correction, here $f_{\rm{e^+}},f_{\rm{n}}$ is the Fermi distribution function of positron and neutron, respectively.

For the neutrino/anti-neutrino scattering by neutrons: $\nu_{\rm{e}} + n\rightarrow \nu_{\rm{e}}+n$ and $\overline{\nu}_{\rm{e}} + n\rightarrow \overline{\nu}_{\rm{e}}+n$,
the resulting scattering coefficient is as follow,
\begin{equation}
\lambda_s=\frac{\sigma_0}{4}\left(\frac{1+3g_{\rm{A}}^2}{4}\right)\left(\frac{p}{m_{\rm{e}}}\right)^2,
\end{equation}

For the neutrino/anti-neutrino scattering by protons:
$\nu_{\rm{e}} + p\rightarrow \nu_{\rm{e}}+p$ and $\overline{\nu}_{\rm{e}} + p\rightarrow \overline{\nu}_{\rm{e}}+p$, the resulting scattering coefficient is written as,
\begin{equation}
\lambda_s=\frac{\sigma_0}{4}\left(4\sin^4 \theta_{\rm{W}} -2\sin^2\theta_{\rm{W}}+\frac{1+3g_{\rm{A}}^2}{4}\right)\left(\frac{p}{m_{\rm{e}}}\right)^2,
\end{equation}
where $\theta_{\rm{W}}$ ($\sin^2{\theta_{\rm{W}}}\sim0.23$) is the Weinberg angle.

\section{Numerical Methods}
In order to simplify the Boltzmann equation to be more compact, we define two new functions $F(z,p,\mu)$  and $G(z,p,\mu)$ :
\begin{eqnarray}
F=f_++f_-,\\
G=f_+-f_-.
\end{eqnarray}
So the Boltzmann equation is transformed to be
\begin{eqnarray}
\mu\frac{\partial{F(z,p,\mu)}}{\partial z}=-(\lambda_a+\lambda_s)G, \\
\mu\frac{\partial{G(z,p,\mu)}}{\partial z}=\lambda_a(2f^{\rm{eq}}-F)+\lambda_s\left(\int_0^1d\mu F - F\right).
\end{eqnarray}
The corresponding boundary conditions $f_+(z=0)=f_-(z=0)$ and $f_-(z=H)=0$ are transformed to be $ G(z=0)=0$ and $F(z=H)=G(z=H)$.
As the first order approximation, we neglect the scattering term in Eq.(10),
then Eqs.(9)-(10) is a set of the standard differential equations with
the boundary conditions.
After obtaining the first order solution $F^1$ and $G^1$ numerically,
we use the first order results to calculate the scattering term in Eq.(10),
after that, the second order solution $F^2$ and $G^2$ are obtained in a similar way.
Repeating this iterative process until a convergent solution is achieved, finally,
the neutrino spectra $F(z,p,\mu)$ and $G(z,p,\mu)$ or equivalently $f_+(z,p,\mu)$ and $f_-(z,p,\mu)$ are obtained.

\section{Neutrino Emitting Spectra}
For simplicity, we consider a characteristic mass density distribution and temperature in the inner part of NDAFs: a distribution of exponent-decreasing mass density, isothermal temperature, iso-electron fraction in the vertical direction, that is,
\begin{equation}
\rho(z)=\rho_c\exp\left(-\frac{z}{z_0}\right)
\end{equation}
where $\rho_c=10^{11}$ g/cm$^3$, $z_0=1$ km, and electron fraction $Y_{\rm{e}}\equiv(n_{\rm{e}}-n_{\rm{e^+}})/n_{\rm{b}}\equiv(n_{\rm{e}}-n_{\rm{e^+}})/(n_{\rm{p}}+n_{\rm{n}})=0.4$ are adopted in our calculation ($n_{\rm{e}}, n_{\rm{e^+}},n_{\rm{b}},n_{\rm{p}},n_{\rm{n}}$ is the number density of electron, positron, baryon, proton, and neutron respectively).

The emitting spectra for neutrino and anti-neutrino $(\frac{p}{kT})^2f(T,p,\mu)$
are shown in Fig.1-2.
In Fig 1, we plot the emitting spectra of neutrinos and anti-neutrinos at the two fixed angles to the vertical direction: 84 degrees (nearly horizontal direction) and 0 degree (precisely vertical direction) under
two typical temperature $T=6$ MeV (left panel) and $T=10$ MeV (right panel). For comparison,
the standard spectrum of the Fermi black body is also shown in the figure.
To investigate the directional dependence of the emitting spectra of neutrinos and anti-neutrinos,
given the neutrino energy ($p=2kT$), the emitting spectra of neutrinos as a function of the emitting direction $\mu=\cos\theta$ are shown in Fig.2. The left and right panels correspond
the results under two typical temperature of the disk ($T=6,10$ MeV).
\begin{figure}
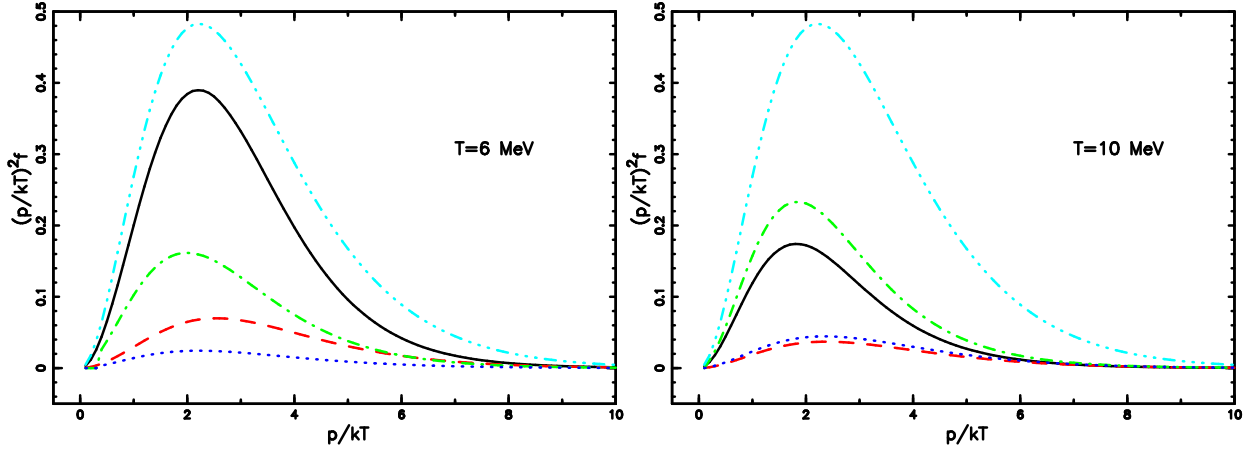

 \centering
 \includegraphics[angle=-90,width=0.5\textwidth]{fig1a}%
 \includegraphics[angle=-90,width=0.5\textwidth]{fig1b}
\caption{Neutrino spectrum $(\frac{p}{kT})^2f(T,p,\mu)$ in two fixed directions at the different temperatures ({\it left panel}: $T=6$MeV, {\it right panel}: $T=10$ MeV). 
The neutrino spectrum in horizontal direction is shown as the solid line, 
the anti-neutrino spectrum in the horizontal direction as the dot-dashed line, 
neutrino spectrum in the vertical direction as the dashed line, 
anti-neutrino spectrum in the vertical direction as the dotted line, 
Fermi black body spectrum as the dot-dot-dashed line,
where the \emph{horizontal direction} here means the direction for $\mu=0.1$ or $\theta=84$ degrees to the vertical direction 
and the \emph{vertical direction} means the direction for 
$\mu=1.0$ or in the vertical direction).}
\end{figure}

\begin{figure}
\includegraphics[width=0.5\textwidth]{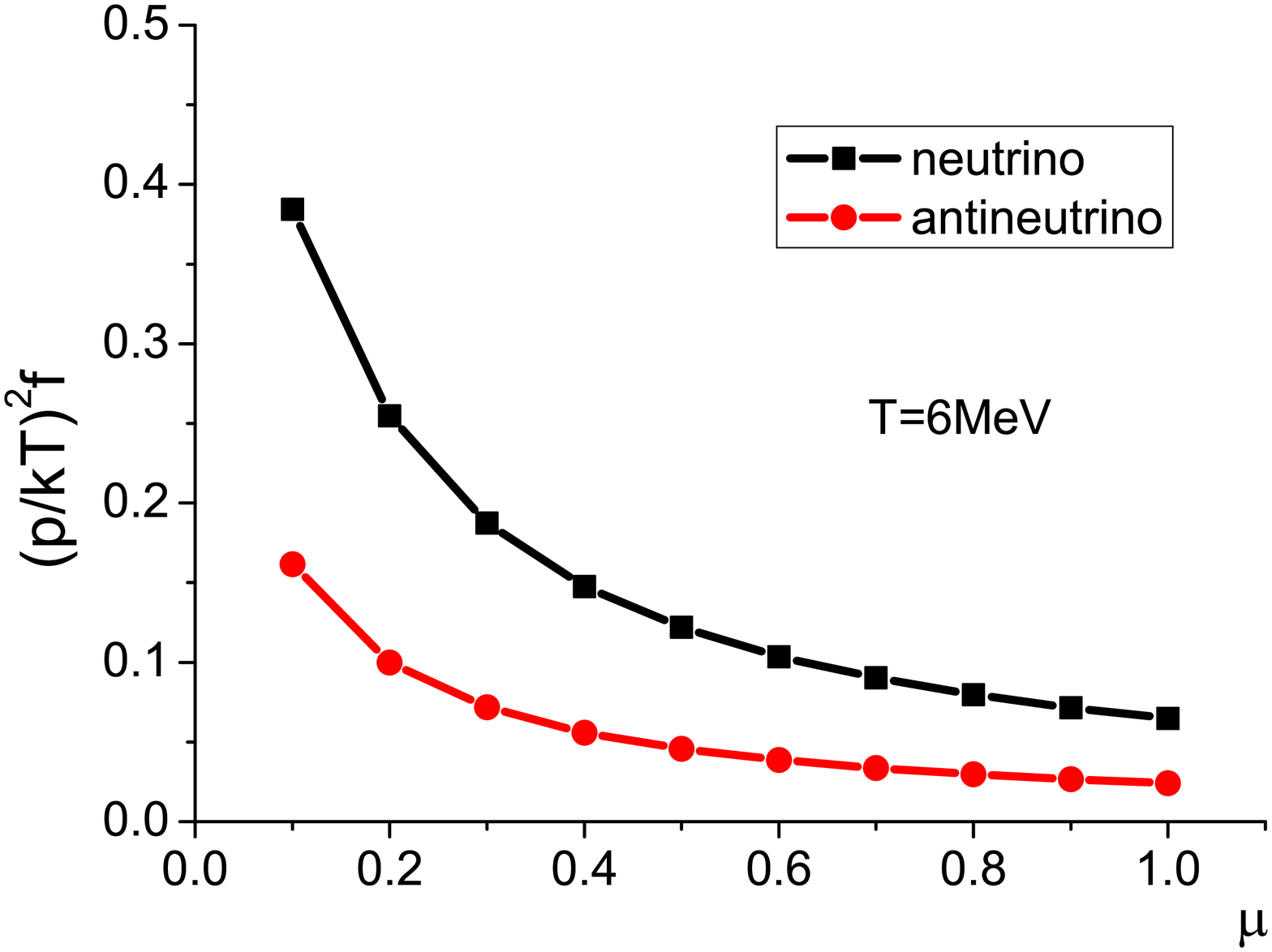}%
\includegraphics[width=0.5\textwidth]{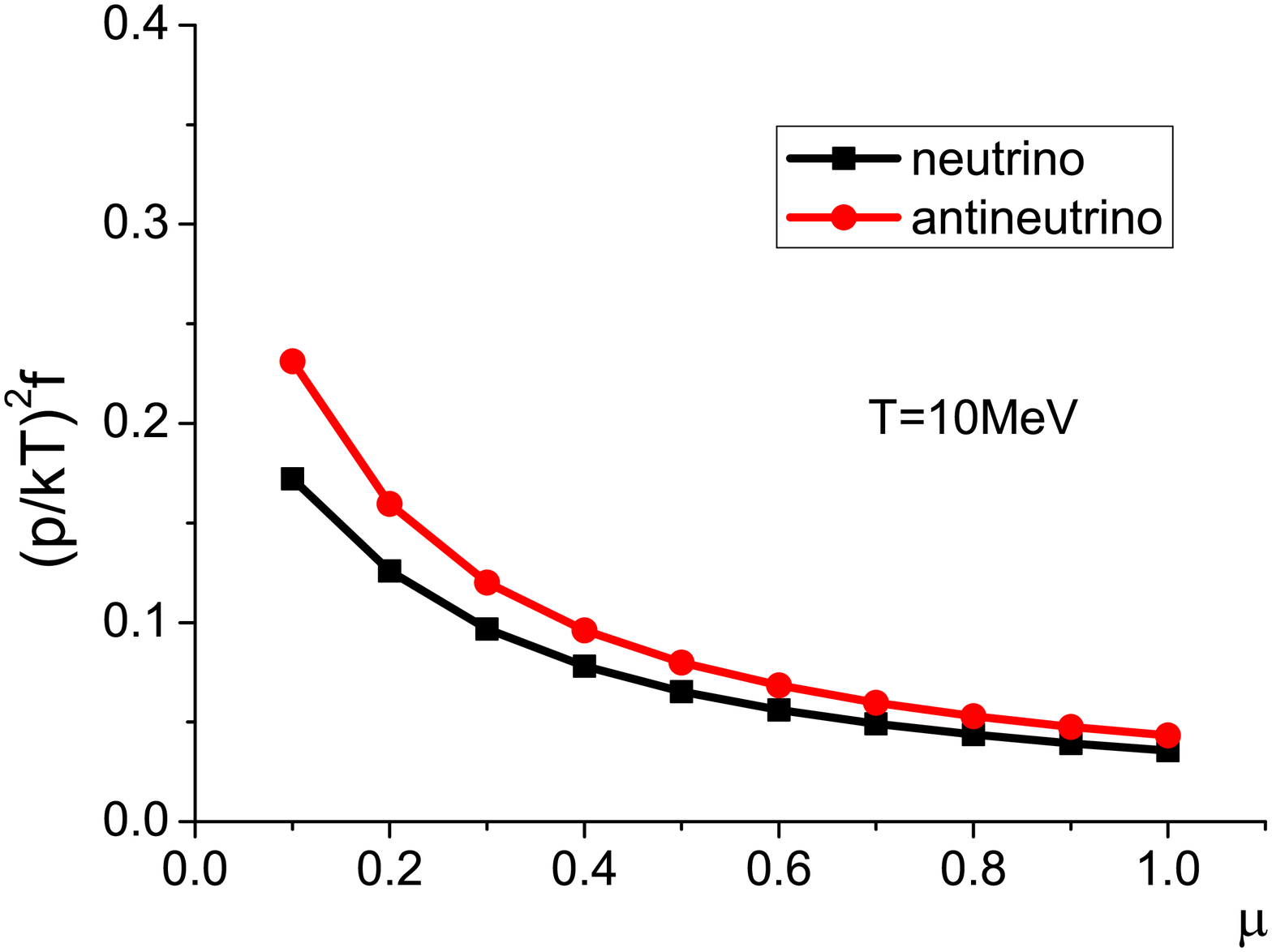}
\caption{Neutrino spectrum $(\frac{p}{kT})^2f(T,p,\mu)$ versus $\mu=\cos\theta$ for a characteristic neutrino energy $p=2kT$
at the different temperatures ({\it left panel}: $T=6$MeV, {\it right panel}: $T=10$ MeV). }
\end{figure}

As shown in Fig 1 and Fig 2, it is evident that the emitting neutrino spectra are
neither the isotropic ones, nor the standard Fermi black body ones.
It is also notified that the neutrino and anti-neutrino emitting spectra are
quite different, especially when the temperature is not very high comparing to
the driving chemical potential $\mu_{\rm{eq}}$, for instance,
the gap between the spectra of neutrinos and that of anti-neutrinos for the case of $T=6$ MeV is much larger than that of $T=10$ MeV.
Both the anisotropic and asymmetric emission of neutrinos and anti-neutrinos will make a difference in the final annihilation rate
of neutrino pairs, as we will discuss in the next section.

\section{Annihilation Rate of Neutrino Pairs}
The annihilation of neutrinos and anti-neutrinos into the electron/positron pairs
above the accretion disk is supposed to be the energy budget of the fireball of SGRBs.
The energy deposition rate by neutrino pairs $\nu_{\rm{e}} , \overline{\nu}_{\rm{e}}$ annihilation is given by\cite{Burrows2006}\cite{Ruffert1997}
\begin{equation}
Q(\nu_{\rm{e}}\overline{\nu}_{\rm{e}} )=\frac{1}{4}\frac{\sigma_0c}{(m_{\rm{e}}c^2)^2(hc)^6}\frac{C_1+C_2}{3}\int_0^{\infty}dp\int_0^{\infty}dp'(p+p')(pp')^3
\int_{4\pi}d\Omega\int_{4\pi}d\Omega'f_{\nu_{\rm{e}}}f_{\overline{\nu}_{\rm{e}}}(1-\cos\Theta)^2
\end{equation}
where $\Theta$ is the angle between neutrino and anti-neutrino beams, and weak coupling constant $C_1+C_2\sim2.34 $.

In this paper, for simplicity, we assume that in the region between $r_{\rm{in}}=10$ km and $ r_{\rm{out}}=200$ km, the mass density is given by Eq.(11), $Y_{\rm{e}}=0.4$ and temperature is constant, then we can precisely calculate the annihilation rate of neutrino pairs, using the obtained
spectra in the above section. The results are listed as follows:
if the disk temperature $T=6$ MeV, the total luminosity of neutrino and anti-neutrino
are $L_{\nu_{\rm{e}}}=2.0 \times 10^{53}$ ergs/s
and $L_{\overline{\nu}_{\rm{e}}}=0.7 \times10^{53}$ ergs/s, respectively.
The annihilation luminosity of neutrino pairs is
$L_{\nu_{\rm{e}}\overline{\nu}_{\rm{e}}}=2.3 \times 10^{50}$ ergs/s,
and the annihilation efficiency $\eta\sim0.1\%$,
here the annihilation efficiency ($\eta$) is defined as $\eta=L_{\nu_{\rm{e}}\overline{\nu}_{\rm{e}}}/(L_{\nu_{\rm{e}}}+L_{\overline{\nu}_{\rm{e}}})$.
If the disk temperature $T=10$ MeV, the total luminosity of neutrino and anti-neutrino are
$L_{\nu_{\rm{e}}}=7.2 \times 10^{53}$ ergs/s
and $L_{\overline{\nu}_{\rm{e}}}=8.8 \times 10^{53}$ ergs/s, respectively.
The annihilation luminosity of neutrino pairs is
$L_{\nu_{\rm{e}}\overline{\nu}_{\rm{e}}}=1.5 \times 10^{52}$ ergs/s,
and the annihilation efficiency $\eta\sim1\%$.

\begin{figure}
\includegraphics[width=0.5\textwidth]{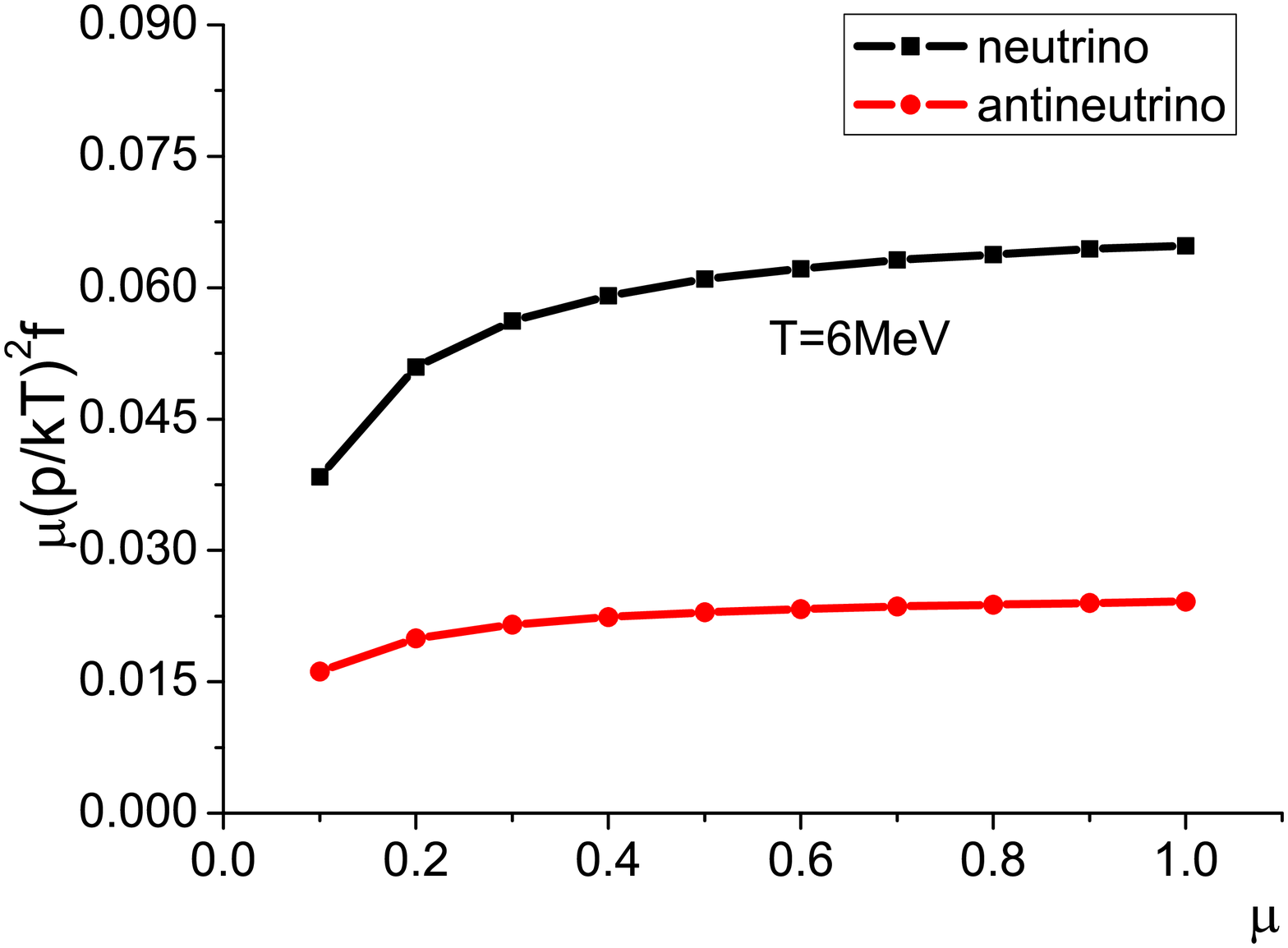}%
\includegraphics[width=0.5\textwidth]{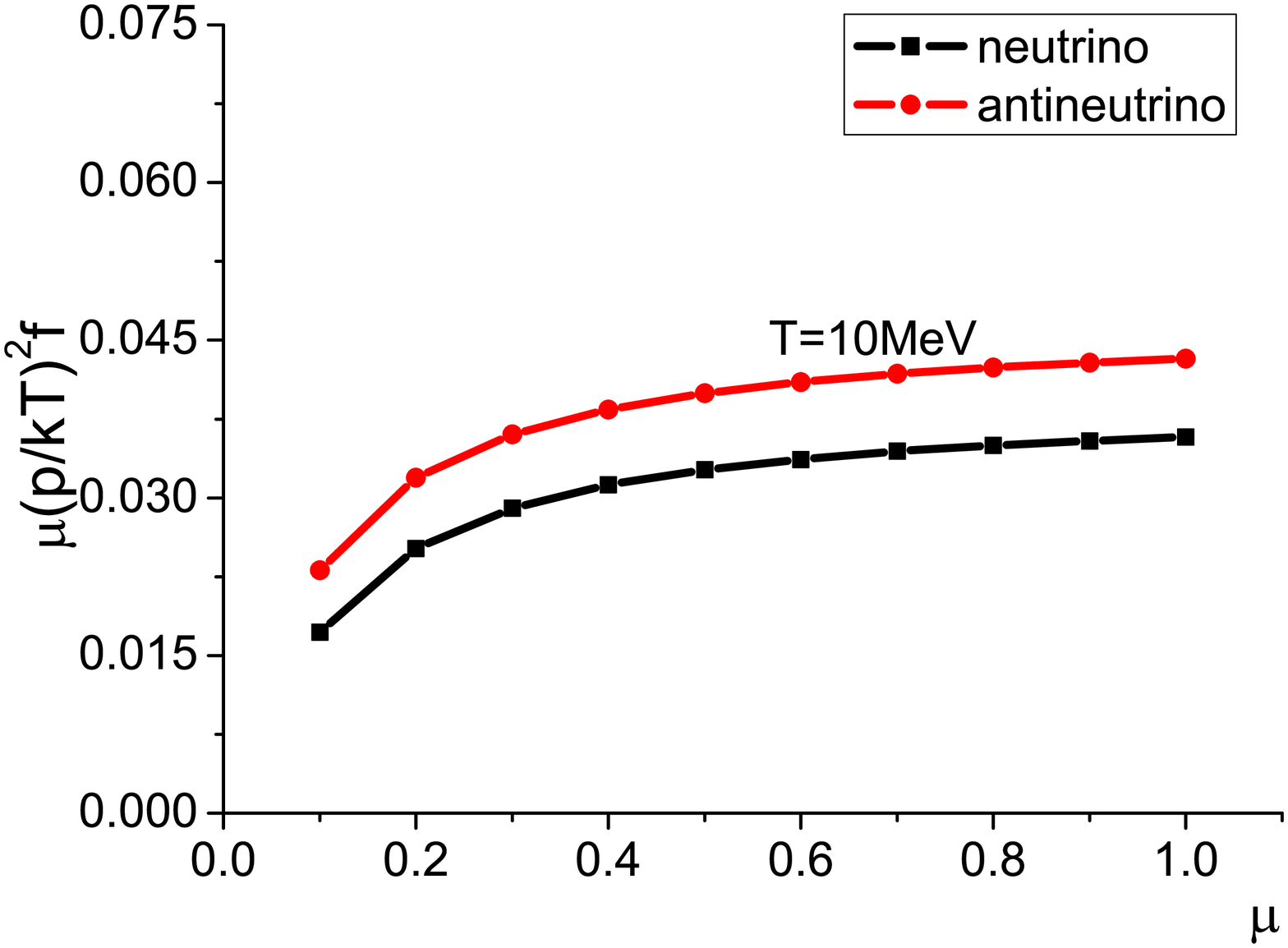}
\caption{Neutrino number flux $\mu(\frac{p}{kT})^2f(T,p,\mu)$ versus $\mu=\cos\theta$ for a characteristic neutrino
energy $p=2kT$ at the different temperatures ({\it left panel}: $T=6$MeV, {\it right panel}: $T=10$ MeV).}
\end{figure}

As discussed above, the neutrino luminosity and their annihilation rate sensitively
depend on the disk temperature, the change of the temperature from 6 MeV to 10 MeV
leads to the two orders magnitude in the final annihilation luminosity!
In our opinion, the reasons are as follows: first, the higher the temperature, the higher
the luminosity of the neutrinos and anti-neutrinos. Second, the higher temperature,
the harder the spectra of neutrinos and anti-neutrinos, which results into
the larger annihilation cross section and the higher annihilation efficiency.
Third, if the temperature is high, the effect of the
driving chemical potential $\mu_{\rm{eq}}$ which
causes the asymmetrical emission of neutrinos and anti-neutrinos is suppressed
(see Fig.2 and Fig.3). The nearly symmetrical emission of neutrinos and anti-neutrinos also
contributes to higher annihilation efficiency.

To support our above arguments,
in Fig.3, we show the neutrino number flux $\mu(\frac{p}{kT})^2f(T,p,\mu)$
(the energy of neutrino and anti-neutrino is taken to be $p=2kT$ )
as a function of the emitting direction $\mu=\cos\theta$, under the different
disk temperature ($T=6$ MeV and $T=10$ MeV). It is obvious that the spectrum of
neutrinos and anti-neutrinos nearly coincide at high temperature ($T=10$ MeV),
while the difference between the spectrum of neutrinos and anti-neutrinos is huge
at low temperature ($T=6$ MeV).

At the end of this section, we will check the validity of the Fermi black body spectrum and the gray body model spectrum.
In Fig 4, we compare the direction averaged neutrino number flux $F_{\rm{\rm{num}}}=(\frac{p}{kT})^2\int_0^1 d\mu f(p,\mu)\mu$ of full Boltzmann equation and Fermi black body spectrum.
It is clearly shown in Fig. 4 that the Fermi black body spectrum overestimates neutrino luminosity greatly (about $5\sim10$ times),
so if the Fermi black body spectrum is adopted to calculate the annihilation luminosity of neutrino pairs  the result will overestimate
about $25\sim100$ times in this case.
Concretely, the Fermi black body spectrum will lead to the following luminosity of
neutrino and anti-neutrino by integrating the neutrino and anti-neutrino spectra \cite{Popham1999}:
\begin{equation}
L=\frac{7}{8}\sigma T^4S
\end{equation}
where $\sigma$ is Stephan-Boltzmann constant, $T$ and $S$ is the temperature and the area of the disk respectively. It is easy to obtain that $L(T=6 \rm{MeV})=1.46\times10^{54}$ ergs/s, and $L(T=10 \rm{MeV})=1.13\times10^{55}$ ergs/s under the condition we considered. 
So the Fermi black body spectrum overestimates the total neutrino and anti-neutrino
luminosity about 5.4 times and the corresponding annihilation rate about 29 times when $T=6$ MeV, and it overestimates the total neutrino and anti-neutrino luminosity 
about 7 times and their corresponding annihilation rate about 49 times.

Similarly, the gray body spectrum will lead to the following luminosity of neutrinos and anti-neutrinos as \cite{DiMatteo2002}\cite{Janiuk2007},
\begin{equation}
L=b\left(\frac{7}{8}\sigma T^4\right)S=\frac{(7/8)\sigma T^4S}{(3/4)(\tau/2+1/\sqrt{3}+1/3\tau_a)}
\end{equation}
where $\tau=\tau_a+\tau_s$ is the sum of scattering optical depth and absorption optical depth, $\tau_a$ is the absorption optical depth and numerically,
\begin{equation}
\tau_s=2.7\times10^{-7}T_{11}^2\rho_{10}H
\end{equation}
\begin{equation}
\tau_a=4.5\times10^{-7}T_{11}^2\rho_{10}H
\end{equation}
where $T_{11}$ is the temperature in unit of $10^{11}$ K, $\rho_{10}$ is the mass density in unit of $10^{10}$ g/cm$^3$ and $H$ is the thickness of disk in unit of cm.
Under the condition we consider ($\rho_{10}=10$, $H=10^5$, and $T_{11}=0.7$ and $1.16$), it is easy to obtain that $\tau(T=6 \rm{MeV})=0.35$, $\tau_a(T=6 \rm{MeV})=0.22$, $b(T=6 \rm{MeV})=0.59$ and  $\tau(T=10 \rm{MeV})=0.96$, $\tau_a(T=10 \rm{MeV})=0.6$, $b(T=10 \rm{MeV})=0.83$. So the gray body spectrum overestimates the total neutrino and anti-neutrino luminosity 
about 3.2 times and annihilation rate about 10 times when $T$=6 MeV, and it overestimates the total neutrino and anti-neutrino luminosity about 5.8 times and their annihilation rate about 34 times when $T$=10 MeV.

In addition, it is commonly assumed that the emission of neutrinos and anti-neutrinos are isotropic and symmetric in calculating their annihilation rate \cite{Popham1999}\cite{Liu2007}. But, as clearly shown in Fig.2, 3 and 4, the emission of neutrinos and
anti-neutrinos are far from isotropic (the intensity in different directions can vary $4\sim5$ times) and far from symmetric (the ratio of the intensity of neutrinos and anti-neutrinos can be $2\sim3$ times). Therefore, it is no doubt that the estimation based on the 
isotropic and symmetric assumption also contribute a non-negligible error in the annihilation rate of neutrino pairs.

Stricktly speaking, we should include the contribution of the annihilation of
the $\mu$ neutrino pairs and $\tau$ neutrino pairs to the total annihilation luminosity, 
but there are two main reasons to consider only the contribution from the electron neutrino
pairs. The first reason is that, as shown in many previous works \citep[e.g.][]{DiMatteo2002,Janiuk2007},
Urca processes ($p+e^- \rightarrow n+\nu_{e}, n+e^+ \rightarrow p+ \bar{\nu}_e$)
dominate the emission of neutrino pairs in the neutrino-cooled disks. Therefore,  
it is reasonable to ignore the luminosity of $\mu$ neutrino and $\tau$ neutrino
pairs from the disk. The second one is that the $\nu_e+\bar{\nu}_e \rightarrow e^- + e^+$ channel 
has the bigger rate than the other two flavors chanels, since the process goes through a 
combination of virtual $Z^0$ and virtual $W^{\pm}$ terms, while the two other species have only 
the only $W^{\pm}$ term.

\begin{figure}
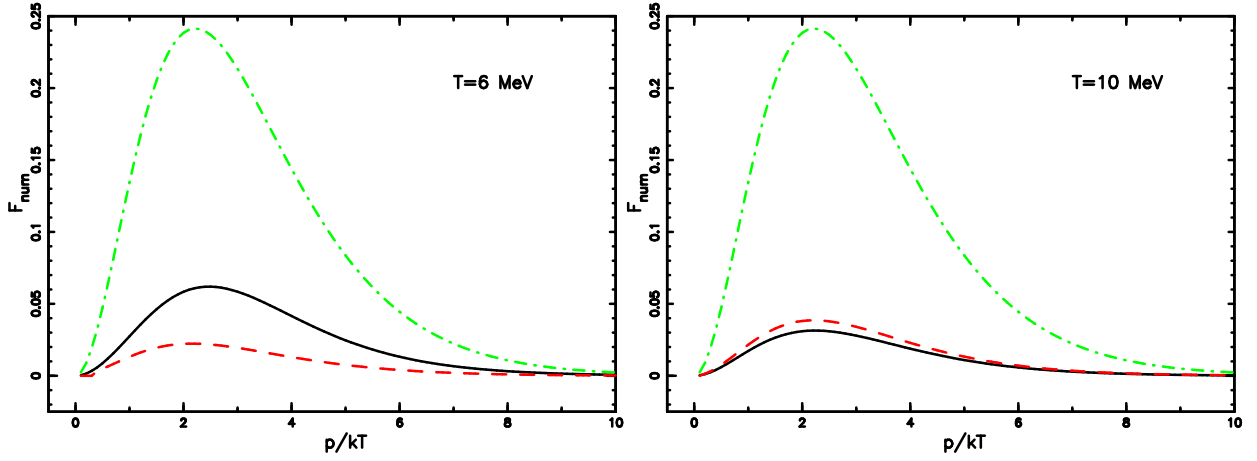

\includegraphics[angle=-90,width=0.5\textwidth]{fig4a}%
\includegraphics[angle=-90,width=0.5\textwidth]{fig4b}
\caption{The direction averaged neutrino/anti-neutrino number flux ($F_{\rm{num}}=(\frac{p}{kT})^2\int_0^1 d\mu f(p,\mu)\mu$)
at the different temperatures ({\it left panel}: $T=6$MeV, {\it right panel}: $T=10$ MeV).
For comparison, the corresponding result for the Fermi black body is also shown: solid line for neutrino, dash line for anti-neutrino, and dot dash line for Fermi black body.}
\end{figure}

\section{Two Stream Approximation}
A simpler way to solve the Boltzmann equation in accretion disk is the two stream approximation\cite{Popham1995} which replaces the full direction dependent distribution by two streams with angle $\cos(\theta)=\pm1/\sqrt3$ to the vertical direction.
In \citet{Sawyer2003}, the author checked the validity of the two stream approximation
of the neutrino transport in accretion disk, and find that
this approximation is good enough.
With this simplification, the Boltzmann equation is written as
\begin{eqnarray}
\frac{1}{\sqrt3}\frac{\partial f_+(z,p)}{\partial z}=\lambda_a[f^{\rm{eq}}(T(z),\mu_{\rm{eq}},p)-f_+(z,p)]+\frac{1}{2}\lambda_s[f_-(z,p)-f_+(z,p)], \\
\frac{1}{\sqrt3}\frac{\partial f_-(z,p)}{\partial z}=-\lambda_a[f^{\rm{eq}}(T(z),\mu_{\rm{eq}},p)-f_-(z,p)]+\frac{1}{2}\lambda_s[f_-(z,p)-f_+(z,p)].
\end{eqnarray}
With the notation we have defined above, under the two stream approximation,
the Boltzmann equation is simplified as follows,
\begin{eqnarray}
\frac{1}{\sqrt3}\frac{\partial{F(z,p,\mu)}}{\partial z}=-(\lambda_a+\lambda_s)G,\\
\frac{1}{\sqrt3}\frac{\partial{G(z,p,\mu)}}{\partial z}= \lambda_a(2f^{\rm{eq}}-F).
\end{eqnarray}
The boundary conditions are the same as the original ones.
The great advantage of the two stream approximation is that the original
Boltzmann equation which is an integral-differential equation is reduced to a
differential equation, which simplifies the original Boltzmann equation greatly.
Obviously, the disadvantage of the two stream approximation is that we lose
the directional information of the emergent spectra of emission.

In Fig 5, we compare the direction-averaged spectra based on the full Boltzmann equation
and the two stream approximation. As seen from Fig.5, the two stream approximation is a rather
good simplification to the Boltzmann equation with high accuracy, if we do not care the concrete angular dependence of the distribution function, e.g. we only care the amount of the
total neutrino number flux or energy flux.
\begin{figure}
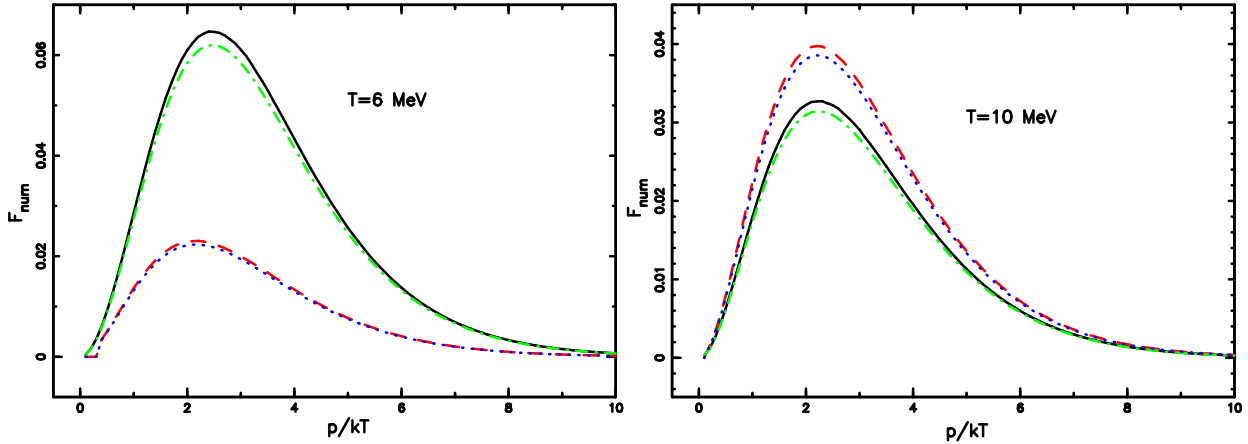

\includegraphics[angle=-90,width=0.5\textwidth]{fig5a}%
\includegraphics[angle=-90,width=0.5\textwidth]{fig5b}
\caption{The direction averaged neutrino/anti-neutrino number flux $F_{\rm{num}}$ resulting from
the full Boltzmann equation and the two stream approximation at the different temperatures ({\it left panel}: $T=6$MeV, {\it right panel}: $T=10$ MeV): solid line and dot dash line are for neutrinos resulting from Boltzmann equation and two stream approximation, dot line and dash line are for anti-neutrinos resulting from Boltzmann equation and two stream approximation correspondingly.}
\end{figure}

\section{Discussion and Conclusion}
In the previous works on the neutrino luminosity and the annihilation rate of neutrino pairs above NDAFs,
the commonly used approximations to determine the luminosity of neutrinos and anti-neutrinos are proved to be
invalid here. Both the Fermi black body spectrum \cite{Popham1999} and the  gray body spectrum \cite{DiMatteo2002}\cite{Janiuk2007} will overestimate the total neutrino and anti-neutrino luminosity and their corresponding annihilation rate greatly under the condition we consider. In addition, other common assumptions on calculating the annihilation rate of neutrino pairs,
such as the emission of neutrinos and anti-neutrinos are isotropic and symmetric \cite{Popham1999}\cite{Liu2007} are proved to be invalid at all, so which will contribute extra
errors in the estimation of the annihilation rate of neutrino pairs.

According to the energy spectra of neutrinos and anti-neutrinos obtained from the Boltzmann equation, we find that the emission of neutrinos and anti-neutrinos are generally asymmetrical. The asymmetrical emission of neutrino and anti-neutrino is an important factor that reduces the annihilation rate of neutrino pairs, and is also responsible for the chemical evolution $Y_{\rm{e}}(t)$ in the disk because the asymmetrical emission of
neutrinos and anti-neutrinos carries the net lepton number flux.
This is an important uncertainty that has not been self-consistently dealt with in the 
previous works,
due to the fact that neither the Fermi black body spectrum nor the simplified neutrino transport model can properly deal with the lepton number flux
carried by the stream of neutrinos and anti-neutrinos.
For instance, the unique property of the atmosphere of neutron stars that the 
lepton number flux and energy flux of the stream of neutrinos and
anti-neutrinos are conserved through the atmosphere, because of its special distribution of
the chemical components\cite{Schinder1982a} \cite{Schinder1982}. 
The presence of the atmosphere of NDAFs similar to the atmosphere of
neutron stars will soften the spectra of the neutrino and anti-neutrino inevitably whatever the incident spectrum is and leave the neutrino luminosity unchanged, thus will suppress the annihilation rate of neutrino pairs above the disk.
But it is well out of the reach of the Fermi black body spectrum or the simplified neutrino transport model.
So only through strictly solving the Boltzmann equation of neutrino transport can we self-consistently deal with the chemical evolution in NDAFs, and then precisely determines
the emitting spectra of neutrino and anti-neutrino, their luminosity and their 
corresponding annihilation rate.

From the strict calculation of the neutrino transport and a coarse assumption about the
distribution of the temperature, mass density and chemical components in the disk, 
we draw the conclusion that a large($r_{\rm{out}}=200$ km) area of very high 
temperature($T\geq10$ MeV) in the neutrino-dominated accretion disk is required 
if the central engine of SGRBs is indeed relevant to the neutrino dominated accretion disk.

Our assumptions on the mass density, the temperature, and the chemical components in the accretion
disk are too simple, and we will do more detailed works on this problem by considering self-consistently the chemical evolution, the energy balance between the
cooling and heating process and the fluid mechanical equilibrium.
As shown in this work, the two stream approximation is accurate enough as long as we do not care
the direction information in the emitting spectra of neutrinos, so will be a prime choice for
dealing with the problem of the chemical evolution and the thermal balance in NDAFs.

\acknowledgements
The authors would like to thank the referee for his/her helpful comments.
This work is partially supported by
National Basic Research Program of China (2009CB824800, 2012CB821800),
the National Natural Science Foundation (11073020, 10733010, 11133005),
and the Fundamental Research Funds for the Central Universities (WK2030220004).

\bibliography{apssamp}

\end{document}